\documentclass[]{revtex4}
\usepackage{graphics}
\usepackage{graphicx}
\usepackage{amsmath}
\usepackage{amssymb}
\def \beq {\begin{equation}}
\def \eeq {\end{equation}}
\def \tr {\rm Tr}
\begin{document}
\title{Lamb shift in radical-ion pairs produces a singlet-triplet energy splitting in photosynthetic reaction centers}
\author{K. M. Vitalis and I. K. Kominis% etc
}                     % Do not remove
\affiliation{Department of Physics, University of Crete, Heraklion 71103, Greece}
\begin{abstract}
Radical-ion pairs, fundamental for understanding photosynthesis and the avian magnetic compass, were recently shown to be biological open quantum systems. We here show that the coupling of the radical-pair's spin degrees of freedom to its decohering vibrational reservoir leads to a shift of the radical-pair's magnetic energy levels. The Lamb shift Hamiltonian is diagonal in the singlet-triplet basis, and results in a singlet-triplet energy splitting physically indistinguishable from an exchange interaction. This could have significant implications for understanding the energy level structure and the dynamics of photosynthetic reaction centers.
\end{abstract}

\maketitle
\section{Introduction}
\label{intro}
Radical-ion pairs \cite{steiner} are biomolecules recently shown \cite{komPRE1,komPRE2,komPRE3,katsop,komCPL1,komCPL2,dellis1,dellis2,cidnp} to exhibit a host of non-trivial quantum effects, providing a strong link between biology and quantum information science, thus further driving the emerging field of quantum biology \cite{JH,briegel,vedral,plenioPRA,sun,horePRL,plenio,plenioPRL,shao,kais}. Radical-ion-pair (RP) reactions are thought to underlie the avian magnetic compass \cite{schulten,ritz,ww,rodgers,mouritsen}, however the primary interest in them stems from their central role in photosynthesis \cite{boxer,matysik}. The photon energy absorbed by the chlorophyll antennae is transformed to an electronic excitation finally reaching the photosynthetic reaction center (RC), and resulting in the transmembrane charge separation essential for biochemical energy production. This charge separation is the end result of a cascade of electron transfers through a series of RPs. So the fundamental understanding of RC dynamics and hence the efficiency of photosynthesis is intimately linked to the understanding of RP reactions.

A new approach for describing spin-selective RP reactions was recently introduced \cite{komPRE1,komPRE2,JH} based on quantum measurement theory. In particular, we showed that spin-selective RP reactions can be understood as an intra-molecule quantum measurement of the RP's electron spin state. This intra-molecule quantum measurement leads to spin decoherence, in particular to singlet-triplet (S-T) dephasing. Now, it is well known that decoherence is one facet of an open quantum system's interaction with its environment. Another is, in principle, a shift in the unperturbed energy levels of the system, generically known as Lamb shift \cite{hornberger,breuer,ivanov,shushin}, first discovered as a splitting of hydrogen's 2S$_{1/2}$-2P$_{1/2}$ levels. 

We will here show that the intramolecule quantum dynamics of radical-ion pairs result in a Lamb shift of their magnetic energy levels, that is, the physical energy levels of the RP are shifted relative to the levels of a "bare" RP, which is an imaginary RP without the decohering vibrational states. It will be shown that, in general, singlet and triplet RP states are shifted by a different amount, resulting in an S-T energy splitting, which is physically indistinguishable from an exchange interaction.

Radical-ion pairs are biomolecular ions with two unpaired electrons and any number of magnetic nuclei, created by a charge transfer from a photo-excited donor-acceptor molecular dyad DA. It is spin dynamics that are of interest in RP reactions, the spin space consisting of the electron and nuclear spins. RPs are usually created from singlet neutral precursors, so their initial state is the singlet electronic spin state denoted by $^{\rm S}{\rm D}^{\bullet +}{\rm A}^{\bullet -}$, where the two dots signify the two unpaired electrons in the donor (+) and acceptor (-) molecular ions. Intra-molecule magnetic interactions, dominated by hyperfine couplings of the RP's electrons to the RP's nuclei, lead to a coherent singlet-triplet mixing of the RP spin state,  $^{\rm S}{\rm D}^{\bullet +}{\rm A}^{\bullet -}\leftrightarrow$ $^{\rm T}{\rm D}^{\bullet +}{\rm A}^{\bullet -}$. While mixing, RP population is lost spin-selectively due to charge recombination taking place at a random instant in time and leading to the neutral reaction products. In the following we do not at all consider recombination reactions, but focus on the state evolution of radical-pairs until the time they recombine.

According to our approach \cite{komPRE1}, the vibrational excitations of the neutral product molecules form a decohering reservoir for the RP's spin evolution. That is,  the coupling of the RP's spin degrees of freedom to the vibrational modes is responsible not only for charge recombination, but it also produces random jumps from the RP state to the reservoir states and back, interrupting the coherent S-T mixing driven by the RP's magnetic Hamiltonian ${\cal H}_{m}$ and leading to S-T dephasing \cite{molin}. The same intramolecule coupling to the vibrational reservoir has yet another consequence:  the physical RP Hamiltonian will be slightly different, shifted from the "bare" Hamiltonian ${\cal H}_{m}$ by the Lamb-shift correction $\delta{\cal H}_{\rm Lamb}$, which we are now going to calculate.

\section{System-Reservoir Interaction}
Consider for the moment just the singlet reservoir, consisting of states with energy $\epsilon_{i}$ (Fig. 1a). These are described by fermionic creation and annihilation operators $a_{i}^{\dagger}$ and $a_{i}$, and the reservoir Hamiltonian is ${\cal H}_{\rm res}=\sum_{i}{\epsilon_{i}a_{i}^{\dagger}a_{i}}$. The fact that we treat a vibrational reservoir with fermionic operators might appear questionable. The reason is that we wish to describe a single occupation of just one of the reservoir states. That is, when the acceptor electron is transferred back to the donor, just one among the quasi-continuous manifold of reservoir states is occupied, and hence this notation is useful to account for this process. This will be evident in the following after we introduce the system-reservoir coupling Hamiltonian in Section 2.1.

If ${\cal H}_{m}$ denotes all magnetic interactions within the radical-pair (hyperfine, Zeeman, etc.), the RP Hamiltonian is $c^{\dagger}c\big(\epsilon_{S}+{\cal H}_{m}\big)$, where $\epsilon_{S}$ is the energy gap of the radical-pair from the neutral precursor (DA) ground state. The operator $c$ describes the occupation of the acceptor's electron site, i.e. $c^{\dagger}c=1$ means the electron is localized at the acceptor and $c^\dagger c=0$ means the electron has moved back to the donor. The sole role of the $c$ operator is to ensure energy conservation for the transitions from the RP state to a reservoir state lying $\epsilon_{S}$ above the ground state.
\subsection{System-reservoir coupling}
The spin degrees of freedom of the radical-pair represent  the open system under consideration. Since the coupling of the RP to the vibrational reservoir states of the neutral product molecule is spin-selective, the amplitude for the transition to one of the singlet reservoir states, $^{\rm S}{\rm D}^{\bullet +}{\rm A}^{\bullet -}\rightarrow{\rm DA}_{i}^{*}$, is proportional to the singlet character of the RP state. Thus the coupling Hamiltonian reads ${\cal V}=\sum_{i}\big(h_{i}+h_{i}^{\dagger}\big)$, where $h_{i}={u_{i}a_{i}^{\dagger}cQ_{S}}$. The operator $Q_{S}$ projects the RP state onto the electron-singlet subspace, while the raising operator $a_{i}^{\dagger}$ produces a single occupation of the $i$-th reservoir level. The transition amplitude $u_i$ will be detailed later. The hermitian conjugate $h_{i}^{\dagger}$ describes the reverse process ${\rm DA}^{*}_{i}\rightarrow$ $^{\rm S}{\rm D}^{\bullet +}{\rm A}^{\bullet -}$. As has been explained in \cite{komPRE1}, a virtual transition $^{\rm S}{\rm D}^{\bullet +}{\rm A}^{\bullet -}\rightarrow{\rm DA}^{*}_{i}$ driven by $h_{i}$, followed by the reverse transition ${\rm DA}^{*}_{i}\rightarrow$ $^{\rm S}{\rm D}^{\bullet +}{\rm A}^{\bullet -}$ driven by $h_{i}^{\dagger}$ produces within 2$^{\rm nd}$-order perturbation theory the fundamental S-T decoherence of radical-pairs. We will now show that these virtual transitions also shift the RP energy levels. 
\begin{figure}
\includegraphics[width=14cm]{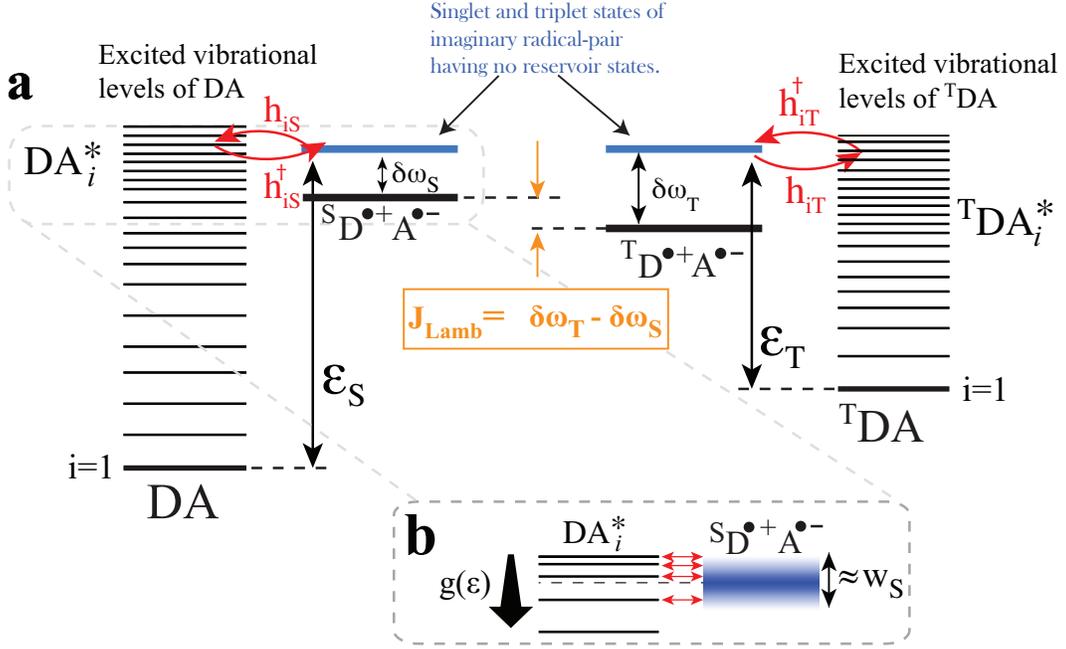}
\caption{Radical-ion pair with singlet and triplet reservoir states. These are the vibrational excitations of the singlet, DA, and triplet, $^{\rm T}$DA, neutral ground states, which form the RP recombination-reaction products. In a bare (unphysical) radical-ion pair without any reservoir states, the S and T states involved in S-T mixing through the magnetic hamiltonian ${\cal H}_{m}$ would be degenerate (blue levels). The presence of reservoir states and the virtual transitions they cause from the radical-pair to them (hamiltonians $h_{iS}$ and $h_{iT}$) and back (hamiltonians $h_{iS}^{\dagger}$ and $h_{iT}^{\dagger}$) do not only damp the S-T coherence, but they also shift the $x$-RP states downwards by $\delta\omega_{x}=k_{x}w_{x}/\epsilon_{0x}$, with $x=S,T$. The resulting S-T energy splitting is $J_{\rm Lamb}=\delta\omega_{T}-\delta\omega_{S}$. (b) The Lamb shift is due to the asymmetry of the Franck-Condon averaged density of states $g_{x}(\epsilon_{x})$ around the finite-lifetime-broadened energy of the $x$-RP.}
\end{figure}
\subsection{Total Hamiltonian and initial state}
RPs are created at $t=0$ in the singlet electron state with a practically zero nuclear spin polarization, hence the initial RP density matrix is $\rho(0)=Q_{S}/\tr\{\it Q_{S}\}$ \cite{note}. The reservoirs states are initially empty (and in the Markovian approximation remain empty throughout the considered evolution), hence the total state of the radical-pair and reservoir is initially $\sigma(0)=\rho(0)\otimes \rho_{E}(0)$, where $\rho_{E}(0)=|0\rangle\langle 0|^{\otimes N}$, with $N$ being the number of singlet reservoir states with energy up to $\epsilon_{S}$. 
The unperturbed Hamiltonian of the total system is ${\cal H}_{0}=c^{\dagger}c(\epsilon_{S}+{\cal H}_{m})+{\cal H}_{\rm res}$, while ${\cal V}$ is the perturbation. The magnetic Hamiltonian ${\cal H}_{m}$ could have been completely omitted and brought back in at the end of the calculation, as its contribution is the simple  unitary evolution $d\rho/dt=-i[{\cal H}_{m},\rho]$. In \cite{komPRE1} the magnetic Hamiltonian was treated as part of the perturbation, again leading to the same result of the ordinary unitary evolution. This is because ${\cal H}_{m}$ effects the state evolution to first order in $dt$, and indeed if we treat it as a perturbation we retrieve the term $d\rho/dt=-i[{\cal H}_{m},\rho]$ within $1^{\rm st}$-order perturbation theory. Decoherence and Lamb shift are derived within $2^{\rm nd}$-order perturbation theory, but interestingly, they also effect reaction dynamics to first order in $dt$.

Finally, the perturbation ${\cal V}$ can be written as a linear combination of system$\otimes$reservoir operators, i.e. ${\cal V}=\sum_{i}S_{i}\otimes E_{i}+{\rm h.c.}$, where $S_{i}=cQ_{S}$ independent of $i$ and $E_{i}=u_{i}a_{i}^{\dagger}$. 
\newline
\subsection{Interaction picture density matrix evolution}
In the interaction picture it is $\tilde{\cal V}=e^{i{\cal H}_{0}t}{\cal V}e^{-i{\cal H}_{0}t}=\sum_{i}(u_{i}e^{i(\epsilon_{i}-\epsilon_{S})t}a_{i}^{\dagger}c\tilde{Q}_{S}+{\rm h.c.})$, where $\tilde{Q}_{S}(t)=e^{i{\cal H}_{m}t}Q_{S}e^{-i{\cal H}_{m}t}$ \cite{note1} and we used the fact that $e^{i\epsilon_{S}tc^{\dagger}c}ce^{-i\epsilon_{S}tc^{\dagger} c}=ce^{-i\epsilon_{S}t}$ and $\tilde{E}_{i}=u_{i}e^{i\epsilon_{i}t}a_{i}^{\dagger}$. The still exact time evolution equation for $\tilde{\sigma}(t)$ is (with $\hbar=1$) $d\tilde{\sigma}/dt=-i[\tilde{\cal V}(t),\sigma(0)]-\int_{0}^{t}{d\tau[\tilde{\cal V}(t),[\tilde{\cal V}(\tau),\tilde{\sigma}(\tau)]]}$. Tracing out the reservoir degrees of freedom after the Born-Markov approximation we arrive at a master equation for the RP density matrix $\tilde{\rho}$:
\beq
{d\tilde{\rho}(t)\over{dt}}=\sum_{ij}\int_{0}^{\infty}\langle E_{i}(0)\tilde{E}_{j}^{\dagger}(\tau)\rangle e^{-i\epsilon_{S}\tau}\Big[\tilde{Q}_{S}(t)\tilde{\rho}(t)\tilde{Q}_{S}^{\dagger}(t-\tau)-\tilde{\rho}(t)\tilde{Q}_{S}^{\dagger}(t-\tau)\tilde{Q}_{S}(t)\Big]d\tau+{\rm h.c.},\label{tilder}
\eeq
where (i) we also traced out the $c$ degrees of freedom as we are interested in the time evolution of the RP state for which $\langle c^{\dagger}c\rangle=1$, and (ii) $\langle A\rangle$ is the expectation value of the reservoir operator $A$ in the state $\rho_{E}(0)$. In particular it is $\langle \tilde{a}_{i}^{\dagger}(\tau)a_{j}(0)\rangle=0$
and $\langle a_{i}(0)\tilde{a}_{j}^{\dagger}(\tau)\rangle=\delta_{ij}e^{i\epsilon_{j}\tau}$. 
Finally, if the eigenvalues and eigenstates of ${\cal H}_{m}$ are denoted by $e_{l}$ and $|e_{l}\rangle$, respectively, then $Q_{S}=\sum_{lm}|lm)$, where $|lm)=\langle e_{l}|Q_{S}|e_{m}\rangle|e_{l}\rangle\langle e_{m}|$, hence in the interaction picture it
will be $\tilde{Q}_{S}(t)=\sum_{lm}e^{-i\omega_{lm}t}|lm)$, where $\omega_{lm}=e_{l}-e_{m}$. Setting the above in Eq. \ref{tilder} we find that 
$d\tilde{\rho}(t)/dt=\sum_{i}\sum_{lm,pq}\Gamma^{i}_{lm,pq}\Big[|lm)\tilde{\rho}(t)(pq|-\tilde{\rho}(t)(pq|lm)\Big]+{\rm h.c.}$, 
where $\Gamma^{i}_{lm,pq}=|u_{i}|^{2}e^{-i(\omega_{lm}-\omega_{pq})t}\int_{0}^{\infty}d\tau e^{i(\epsilon_{i}-\epsilon_{S}-\omega_{pq})\tau}$. 
\subsection{Master equation for unreacted radical pairs}
To go back to the Schr\"{o}dinger picture we replace $\tilde{\rho}$ in the LHS of the above equation for $d\tilde{\rho}/dt$ by $e^{i{\cal H}_{m}t}\rho e^{-i{\cal H}_{m}t}$ and then multiply both sides from the left by $e^{-i{\cal H}_{m}t}$ and from the right by 
$e^{i{\cal H}_{m}t}$. The LHS will lead to $d\rho/dt+i[{\cal H}_{m},\rho]$.  To evaluate the RHS we note that
(i) $e^{-i{\cal H}_{m}t}|lm)\tilde{\rho}(pq|e^{i{\cal H}_{m}t}=e^{i(\omega_{lm}-\omega_{pq})t}|lm)\rho(pq|$,  (ii) since $\sum_{lm,pq}(pq|lm)=Q_{S}^{2}=Q_{S}$ (${\rm Q_S}$ is a projector), it easily follows that $\sum_{lm,pq}e^{-i{\cal H}_{m}t}\tilde{\rho} (pq|lm)e^{i{\cal H}_{m}t}=e^{i(\omega_{lm}-\omega_{pq})t}\rho Q_{S}$ and (iii) in the integral over $\tau$ in the expression for $\Gamma_{lm,pq}^{i}$ we omit $\omega_{pq}\tau$ in the phasor since $\omega_{pq}\ll\epsilon_{i}-\epsilon_{S}$ \cite{note2}. The master equation then
becomes 
\beq
d\rho/dt=-i[{\cal H}_{m},\rho]+\Gamma(Q_{S}\rho Q_{S}-\rho Q_{S})+{\rm h.c.},\label{drdt}
\eeq
where $\Gamma=\sum_{j}\Gamma_{j}$ and $\Gamma_{j}=|u_{j}|^2\int_{0}^{\infty}d\tau e^{i(\epsilon_{j}-\epsilon_{S})\tau}$.

The amplitude $u_j$ is composed of an electronic matrix element and a nuclear overlap matrix element, $u_j=v_{j}\chi_{j}$ \cite{jortnerET}. We consider the former to be independent of $j$ in the vicinity of $\epsilon_{S}$, $v_{j}=v$, and introduce the Franck-Condon averaged density of states $g_{S}(\epsilon)=|\chi(\epsilon)|^2d(\epsilon)$, which takes into account both $\chi(\epsilon)$, the nuclear wave function overlap integral, and $d(\epsilon)$, the density of vibrational states at the energy $\epsilon$. The discrete sum $\sum_{j}\Gamma_{j}$ is thus approximated by the integral $|v|^{2}\int d\epsilon g_{S}(\epsilon)\int_{0}^{\infty}d\tau e^{i(\epsilon-\epsilon_{S})\tau}$.
Introducing the detuning $\Delta=\epsilon-\epsilon_{S}$ and noting the well-known relation for Heaviside's function Fourier transform, $\int_{0}^{\infty}e^{-i\omega t}dt=\pi\delta(\omega)+\mathbb{P} {1\over i\omega}$, we find 
\beq
\Gamma\equiv\gamma_{R}-i\gamma_{I}=|v|^{2}\int d\Delta\Big[\pi\delta(\Delta)+\mathbb{P}{1\over {i\Delta}}\Big]g_{S}(\Delta+\epsilon_{S})\label{gamma}
\eeq
\section{Lamb shift in radical-ion pairs is physically equivalent to a spin-exchange interaction}
When the real part of \eqref{gamma} is inserted into \eqref{drdt} we will arrive at the Lindblad description of S-T decoherence already discussed in \cite{komPRE1}. Using the imaginary part of \eqref{gamma} in \eqref{drdt} leads to the Lamb shift Hamiltonian. Before proceeding to the latter, we note that the real part of \eqref{gamma}, $\gamma_{R}=\pi|v|^{2}g_{S}(\epsilon_{S})$, is nothing else than half the singlet recombination rate, $k_{S}/2$. Indeed, recombination of radical-pairs proceeds within 1$^{\rm st}$-order perturbation theory by a real transition to a reservoir state, followed by another real transition (decay) of the reservoir state to the radical-pair's ground state DA. The latter happens very fast, e.g. at ps timescales, so the rate limiting process is the former. Using Fermi's golden rule we immediately find that the recombination rate will be $k_{S}=2\pi |\langle f|{\cal V}|i\rangle|^{2}d(\epsilon_{S})=2\pi|v|^2g_{S}(\epsilon_{S})$, where as initial state $|i\rangle$ we chose a pure singlet state of the radical-pair, and as a final state $|f\rangle$ one among the near-resonant and quasi-continuum reservoir states described by the density of states $d(E)$.

To find the imaginary part $\gamma_{I}$ we expand $g_{S}(\Delta+\epsilon_{S})\approx g_{S}(\epsilon_{S})+g_{S}'(\epsilon_{S})\Delta$. The $\Delta$-integration range is determined by the RP's energy uncertainty. Since the singlet radical-pair has a finite lifetime $\tau_{S}$, it's energy level will be broadened by $w_{S}\approx 1/\tau_{S}$ according to Heisenberg's energy-time uncertainty. Since $\mathbb{P}(1/\Delta)$ is an odd function of $\Delta$, only the second term of the previous  expansion will survive the integration. The result will be
$\gamma_{I}=|v|^{2}g_{S}'(\epsilon_{S})w_{S}$. Putting everything together (also using the hermitian conjugate term in \eqref{drdt}), and repeating the above calculation for the triplet reservoir states, we find that the density matrix of unreacted radical pairs evolves according to
$d\rho/dt=-i[{\cal H}_{m}+\delta{\cal H}_{\rm Lamb},\rho]+{\cal L}(\rho)$, where the Lamb shift Hamiltonian finally reads
\beq
\delta{\cal H}_{\rm Lamb}=\delta\omega_{S}Q_{S}+\delta\omega_{T}Q_{T},
\eeq
where
\beq
\delta\omega_{x}={1\over {2\pi}}{{g'_{x}(\epsilon_{x})}\over {g_{x}(\epsilon_{x})}}k_{x}w_{x},~~~~x=S,T\label{shift}
\eeq
For completeness we reiterate that ${\cal L}(\rho)=-{k_{S}\over 2}(Q_{S}\rho+\rho Q_{S}-2Q_{S}\rho Q_{S})-{k_{T}\over 2}(Q_{T}\rho+\rho Q_{T}-2Q_{T}\rho Q_{T})$ is the aforementioned S-T dephasing term \cite{komPRE1}.

The physical interpretation of $\delta{\cal H}_{\rm Lamb}$ is based on these two points: (i) the RP energy levels acquire a broadening due to their finite lifetime, and (ii) the Franck-Condon averaged density of states $g(\epsilon)$ (for simplicity we will henceforth omit the $S$ or $T$ index of the function $g$ and its argument $\epsilon$) has a steep $\epsilon$-dependence, in fact $dg(\epsilon)/d\epsilon<0$ \cite{jortner1,jortner2}. Hence when the RP makes a virtual transition to a reservoir level, it momentarily acquires an energy which is smaller, on average, than the bare RP energy. The physical energy levels of the singlet and triplet RP are thus shifted downwards by $\delta{\cal H}_{\rm Lamb}$ with respect to the bare RP. 

In the general case when $\delta\omega_{S}\neq\delta\omega_{T}$, the Lamb shift will lead to an S-T {\it energy splitting}, as if there was an exchange coupling of the form $J\mathbf{s}_{1}\cdot\mathbf{s}_{2}$, which is known to produce an S-T energy splitting of $J$ (the triplet being higher in energy for $J>0$). In other words, the differential Lamb shift of singlet and triplet RP states, which is inherent in the RP due to the interaction with its intra-molecule vibrational reservoir, would physically look as an exchange interaction having $J_{\rm Lamb}=\delta\omega_{T}-\delta\omega_{S}$. Put differently, if an exchange interaction with coupling $J$ actually exists in the RP, the physically observed S-T energy splitting will not be $J$ but $J'=J+J_{\rm Lamb}$. 

\begin{figure}
\includegraphics[width=10cm]{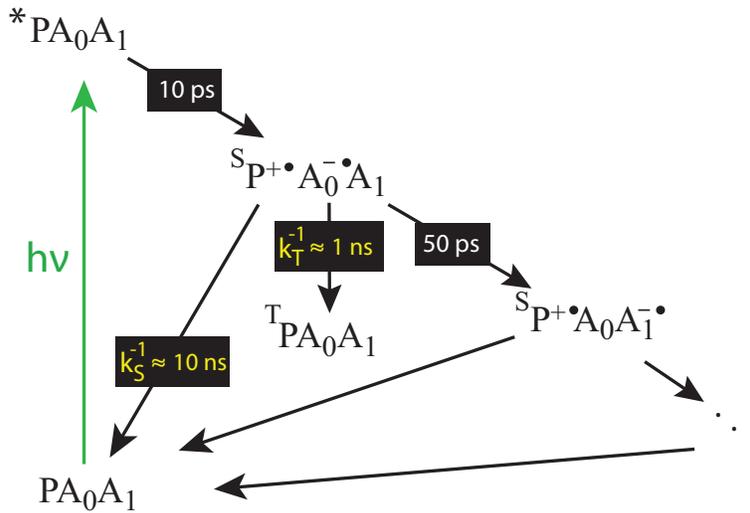}
\caption{First stages of charge separation in the photosynthetic reaction center of Photosystem I. The goal of the RC is to efficiently create the final charge-separated state (after the dots in the lower right part of the figure) starting from the photo-excited  $^{*}{\rm P}{\rm A_0}{\rm A_1}$ and going through a series of electron-transfers to intermediate radical-ion pairs. P and A0 are chlorophyll molecules, and A1 phylloquinone. The lifetime of the intermediate radical-pair $^{\rm S}{\rm P}^{\bullet +}{\rm A_0}^{\bullet -}{\rm A_1}$ is not dominated by the recombination rates $k_S$ and $k_T$, but by the much faster electron transfer (50 ps timescale) to the following charge separated state.}
\end{figure}

The sign of $J_{\rm Lamb}$ is determined by the sign of $\delta\omega_{S}$ and $\delta\omega_{T}$, and their relative size. Since $g'(\epsilon)<0$, it is $\delta\omega_{S},\delta\omega_{T}<0$. Hence the sign of $J_{\rm Lamb}$ will be the sign of $|\delta\omega_{S}|-|\delta\omega_{T}|$. How large is $|J_{\rm Lamb}|$ in realistic cases? To this end we have to first evaluate the typical value of $g'(\epsilon)/g(\epsilon)$. It is known that $g(\epsilon)$ can be locally approximated \cite{jortner1,jortner2} by an exponential, $g(\epsilon)\propto e^{-\epsilon/\epsilon_{0}}$, where $\epsilon_{0}\approx 200-700~{\rm cm}^{-1}\approx 10^{13}~{\rm Hz}$, hence $|g'(\epsilon)/g(\epsilon)|\approx 1/\epsilon_{0}$. Thus, apart from constants of order unity, the shifts will be of order 
\beq
\delta\omega\approx {{kw}\over \epsilon_{0}}
\eeq
For isolated radical-pairs the singlet and triplet lifetimes are determined by the spin-selective charge-recombination rates, i.e. 
$w=k$, hence $\delta\omega=k^{2}/\epsilon_{0}$.  For a typical recombination rate of $k\approx 10^{9}~{\rm s}^{-1}$ it follows that $\delta\omega\approx 10^{5}~{\rm s}^{-1}$, which expressed in magnetic-field units is $\delta\omega\approx 5~{\rm mG}$.
The splitting $J_{\rm Lamb}$, which will result if $k_{S}\neq k_{T}$ and/or $\epsilon_{0S}\neq\epsilon_{0T}$ will be of the same order. Although small, such splittings can produce interesting low-field level-crossing effects to be addressed elsewhere.   
\section{Radical-Ion-Pair Lamb Shift in Photosynthetic Reaction Centers}
The considered effect becomes much larger for radical-pairs participating in the charge-separation pathway in photosynthetic reaction centers, as shown in Fig. 2. Photosynthetic RCs exhibit a cascade of electron-transfer steps until the stable charge-separated state is produced. In each of those steps a different RP is formed, and its inverse lifetimes $w_{S}$ and $w_T$ are also influenced, and in cases dominated, by the electron-transfer rates to the following step. For example, the lifetime of the singlet RP $^{\rm S}{\rm P}^{\bullet +}{\rm A_0}^{\bullet -}{\rm A_1}$ shown in Fig.2 is not dominated by the singlet or triplet recombination time of about 10 ns and 1 ns, respectively, but by the electron transfer to the next-stage RP, $^{\rm S}{\rm P}^{\bullet +}{\rm A_0}{\rm A_1}^{\bullet -}$, taking place in about $w^{-1}\approx 50~{\rm ps}$. In this case, there will be two different shifts, one stemming from reservoir states of the singlet and triplet neutral products, $^{\rm S}{\rm PA_0A_1}$ and $^{\rm T}{\rm PA_0A_1}$, respectively, and one stemming from the reservoir states of the next-stage RP, ${\rm P}^{\bullet +}{\rm A_0}{\rm A_1}^{\bullet -}$. For the former it will be $\delta\omega_{x}\approx k_{x}w/\epsilon_{0}$, where we took $\epsilon_{0S}\approx\epsilon_{0T}\approx\epsilon_{0}$. Since $k_{T}\gg k_{S}$, it will   be $\delta\omega_{T}\gg\delta\omega_{S}$ and $J_{\rm Lamb}\approx\delta\omega_{T}\approx -0.1~{\rm G}$. For the latter the shifts will be on the order of $\delta\omega\approx-w^{2}/\epsilon_{0}\approx -2~{\rm G}$, which is a significant shift. Since the electron transfer rate to the next-stage RP is not spin-dependent ($w_{S}=w_{T}=w$) it would appear that both the singlet and the triplet RP would be shifted by $\delta\omega$. However, the interplay of these shifts with RC dynamics is more involved and will be explored elsewhere. 
\section{Discussion}
We finally comment on our perturbative derivation of the Lamb shift Hamiltonian. In the application of the Lamb shift expression \eqref{shift} in photosynthetic RCs we used as input the short RP lifetime $w^{-1}\approx 50~{\rm ps}$, so the use of ${\cal V}$ as a perturbation might appear questionable. In other words, the relevant rate $w$ related to the perturbation ${\cal V}$ is much larger than the typical frequency scale of the unperturbed magnetic Hamiltonian ${\cal H}_{m}$. To alleviate such a concern we first note that there is a similar example in NMR \cite{nmr}, where spin-lattice relaxation theory is equally applicable at low magnetic fields where the typical Larmor frequency is smaller than the relaxation rate. We secondly note that the calculated magnitude of the shift is of the same order as ${\cal H}_m$, so at least the use of perturbation theory produces a consistent result. Finally, the actual reason why our perturbative treatment does not pose a problem is that the high reaction rate $w$ strongly depends on the high density of states of the reservoir, i.e. each individual term in the expansion ${\cal V}=\sum_{i}S_{i}\otimes E_{i}$ can indeed be considered and treated as a perturbation, while the final rates depends on the combined action of all those terms. Indeed, one of the main starting assumptions of all calculations such as the one presented in \cite{jortner1} (see e.g. equation II.1 of \cite{jortner1}) is that the number of reservoir states within the lifetime-broadened width $w$ of the RP is much larger than one. This is expected since the energy gap $\epsilon_{S}$ relevant for the reservoir density of states is about two orders of magnitude higher than the typical vibrational frequency. 
\section{Conclusions}
Concluding, we have analyzed the complete effect of the continuous quantum measurement taking place in RPs as a result of their coupling to the intramolecule vibrational reservoir. Besides the spin decoherence that was described in our earlier work, we  here considered the shift this quantum measurement brings about to the RP energy levels. This shift can have non-negligible values in photosynthetic reaction centers. Since this shift is equivalent to an exchange interaction, which is known to suppress singlet-triplet mixing and thus directly affect RP spin dynamics, it will be interesting to examine the effect of such shifts in the dynamics of RCs. 
 
\acknowledgements This research has been co-financed by the European Union (European Social Fund - ESF) and Greek national funds through the Operational Program "Education and Lifelong Learning" of the National Strategic Reference Framework (NSRF) - Research Funding Program THALIS, and by the European Union's Seventh Framework Programme FP7-REGPOT-2012-2013-1 under grant agreement 316165. I.K.K. would like to acknowledge discussions with Prof. Steven Boxer and Prof. Ulrich Steiner.


\begin{thebibliography}{}
\bibitem{note}
The dimension of the RP's density matrix is $d=4\Pi_{j=1}^{n}(2I_{j}+1)$, where 4 is the spin multiplicity of the 2 electrons and the rest is the nuclear spin multiplicity of $n$ magnetic nuclei with spin $I_1$, $I_2$, ..., $I_n$. The
singlet and triplet projection operators $Q_{S}=1/4-\mathbf{s}_{1}\cdot\mathbf{s}_{2}$ and $Q_T=3/4+\mathbf{s}_{1}\cdot\mathbf{s}_{2}$ are d-dimensional matrices . The initial density matrix $\rho(0)=Q_{S}/\tr\{Q_{S}\}$ is properly normalized ($\tr\{\rho(0)\}=1$) and it describes a singlet electronic spin state ($\tr\{\rho(0) {\rm Q_S}\}=1$) with completely unpolarized nuclear spins, i.e. $\tr\{Q_{S}I_{\alpha,j}\}=0$ for components $\alpha=x,y,z$ of the $j$-th nuclear spin.
\bibitem{note1}
The projector $Q_{S}$ does not in general commute with the magnetic Hamiltonian ${\cal H}_{m}$.
\bibitem{note2}
Obviously $\epsilon_{S}\gg \omega_{pq}$, where $\omega_{pq}$ is a typical frequency scale of the magnetic hamiltonian ${\cal H}_{m}$, since $\epsilon_{S}\approx 1~{\rm eV}\approx 10^{14}~{\rm Hz}$ is of electrostatic origin, whereas $ \omega_{pq}\approx  10^{7}~{\rm Hz}$ is determined by typical hyperfine couplings of about 10 G. The frequency  $\omega_{pq}$ is also much smaller than $\epsilon_{i}-\epsilon_{S}$, since the latter is on the order of the detuning $\Delta$, which is on the order of $k_{S}\approx 10^9~{\rm s}^{-1}$ or even higher.
\end{thebibliography}
\end{document}